\documentclass[aps,pre,showpacs,onecolumn]{revtex4}
\usepackage{graphics}
\usepackage[dvips]{color}
\usepackage[T1]{fontenc}
\usepackage[latin1]{inputenc}
\usepackage{graphicx}
\usepackage{amssymb}
\usepackage{rotating}
\usepackage{epsfig}

\input epsf.sty

\newcommand{\sts}{\scriptsize}

\newcommand{\mb}{\mbox}

\begin{document}

\title{Influence of  correlations on  molecular recognition}
\author{Hans Behringer, Friederike Schmid}

\affiliation{Fakult\"at f\"ur Physik, Universit\"at Bielefeld, D -- 33615
Bielefeld, Germany}

\begin{abstract}

  \noindent The influence of the patchiness and correlations in the
  distribution of hydrophobic and polar residues at the interface
  between two rigid biomolecules on their recognition ability is
  investigated in idealised coarse-grained lattice models.  A general
  two-stage approach is utilised where an ensemble of probe molecules
  is designed first and the recognition ability of the probe ensemble
  is related to the free energy of association with both the target
  molecule and a different rival molecule in a second step. The
  influence of correlation effects are investigated using numerical
  Monte Carlo techniques and mean field methods. Correlations lead to
  different optimum characteristic lengths of the hydrophobic and
  polar patches for the mutual design of the two biomolecules on the
  one hand and their recognition ability in the presence of other
  molecules on the other hand.

\end{abstract}


\maketitle

\section{Introduction}

An understanding of the basic principles of biomolecular recognition,
that is the ability of a biomolecule to interact selectively with
another molecule in the presence of structurally similar rival
molecules, is not only important from a scientific point of view but
also opens up a wide field of potential biotechnological applications
\cite{Alberts_1994,Kleanthous_2000,Peppas_2002}.  The recognition
process itself is governed by a complex interplay of non-covalent
interactions such as salt bridges, hydrogen bonds, van der Waals and
hydrophobic interactions.  The typical intrinsic energy contribution
of such an interaction is of the order of 1-2 kcal/mol and is thus
only slightly larger than the thermal energy
$k_{\sts\mb{B}}T_{\sts\mb{room}} = 0.62$ kcal/mol at room temperature
\cite{Delaage_1991,Sneppen_2005}.  In order to stabilise a complex of
two proteins over a time long enough to ensure its biological function,
many favourable interactions have to be established to overcome the
entropic cost of the formation of the complex.  Therefore, the two
molecules have to complement each other at the common interface with
respect to shape and interaction partners \cite{Pauling_1940}.  This
principle of complementarity is closely related to the lock-and-key
view of rigid protein-protein recognition \cite{Fischer_1894}. 

Molecular recognition results from an interplay of numerous competing
and cooperating factors.  Apart from the scenario of recognition
between rigid proteins, recognition processes where at least one of
the biomolecules undergoes conformational changes are also numerous in
nature. Such recognition processes are described by the induced fit
scheme \cite{Koshland_1958}.  To understand the recognition process in
full, one not only needs to consider the stability of a single
specific complex, but also the encounter of the two biomolecules in
the heterogeneous environment of the cell. For example, long-range
electrostatic interactions are believed to pre-orient the biomolecules
so that the probability of an encounter of the complementary patches
on the two molecules upon collision is increased
\cite{Kleanthous_2000,Wodak_2003}. Another critical aspect is the
competition due to the simultaneous presence of different molecules.
The more the binding free energy between complementary biomolecules
differ from the binding free energy to other molecules the lower is
the risk of misrecognition.

The recognition problem of two biomolecules shows up in different
disguises in nature. To gain insight into this problem different
approaches can be adopted. A detailed modelling (often on an atomistic
level) of the biomolecules that form a complex gives many insights
into the actual binding process between two specific biomolecules.  In
drug design docking methods allow the identification of the drug
molecule with the optimum binding affinity for a known biomolecule. A
second way to investigate the problem of molecular recognition is the
use of coarse-grained models. The study of idealised coarse-grained
and hence abstract generic models with methods from statistical
physics seems to be particularly adequate for an understanding of the
basic common physical mechanisms that govern different recognition
processes in the heterogeneous environment of a cell. The
coarse-graining approach is based on a reduction to the most relevant
degrees of freedom for molecular recognition which helps to abstract
from complications due to the intricate interplay of the involved
types of interactions so that the generic features nature exploits for
recognition can be identified \cite{Behringer_etal_2007}. This
approach has been adopted in the literature to analyse various aspects
of biomolecular binding and recognition for (almost) rigid and
flexible biomolecules in idealised model systems
\cite{Lancet_1993,Janin_1996,Janin_1997,Rosenwald_2002,Wang_2003,Bogner_2004,Bernauer_2005,
  Bachmann_2006,Lukatsky_2006}.

On popular approach to study the basic principles of molecular
recognition consists in investigating the adsorption of heteropolymers
on patterned surfaces. Biomolecular recognition is then viewed in a
first approximation as the adsorption of a biopolymer on the surface
of another biopolymer.  One major aspect addressed in this context
deals with the question, whether or not length scale matching on the
two polymers favours adsorption
\cite{Muthukumar_1995,Bratko_1997,Golumbfskie_etal_1999,Chakraborty_2001,
  Polotsky_2004a,Polotsky_2004b, Ziebarth_2008}.  Generally speaking
it was found that the adsorption properties depend on the involved
types of correlations and that statistically structured surfaces (be
it correlated or anticorrelated ones) have an enhanced affinity
towards similarly structured chains although an exact matching of the
corresponding correlation lengths is not necessary. The adsorption is
followed by a second freezing transition where the flexible chain
adjusts to the pattern of the surface which necessitates a more
precise matching of the correlation lengths.  Bogner et al.
\cite{Bogner_2004} also addressed the role of correlations and found
that biomolecular binding seems to be strongly influenced by small
scale structures suggesting that local structure elements are
particularly important for molecular recognition.

The present study is in some sense complementary to those works. We
investigate the influence of correlation effects on molecular
recognition within coarse-grained models that are specifically
designed to model the recognition between almost rigid proteins. In
particular we focus on the role of the presence of competing rival
molecules on the recognition characteristics. In our model
correlations appear in the distribution of hydrophobic and polar
residues on the surface of a biomolecule. These correlations result in
extended patches of several hydrophobic and polar residues on the
surface of the protein. The patterns of the actual target molecule and the rival
molecules thereby exhibit the same characteristic correlation lengths.
We then address the question about the optimum correlation length of
the biomolecule that is supposed to recognise the target. All in all
our analysis shows that a matching of the patterns on the surfaces is
necessary to a certain degree in order to get optimum selectivity.
However, the precise way how the correlation lengths fit to each other depends
on whether or not rival molecules are present, that is whether the
isolated binding process or whether the actual recognition process
with rival molecules present is considered. We note also that in a
recent study the effect of correlations that stem from the density of
atoms on the surface of a biomolecule was considered in the context of
connected proteins in protein interaction networks
\cite{Lukatsky_2008}.

The present article is organised in the following way. In the next
section our general approach to biomolecular recognition of two rigid
proteins in the presence of rival molecules is briefly sketched (for a
more detailed account, see \cite{Behringer_etal_2006,
  Behringer_etal_2007b}).  In the subsequent section \ref{kap:korrel}
we discuss how correlations in the distribution of hydrophobic and
polar residues can be incorporated into the model. In sections
\ref{kap:unkoop} and \ref{kap:domkoop} we then investigate the
influence of sequence correlations on molecular recognition by using
Monte Carlo techniques and mean field approximations.

\section{Model and general approach to molecular recognition}

In this work we use coarse-grained idealised model systems to
investigate the recognition of two biomolecules.  Coarse-grained model
systems contain a limited number of degrees of freedom and hence the
recognition problem in its various disguises can not be captured in
its full scope. We limit our investigations to recognition processes
that belong to the scenario of rigid protein-protein recognition and
consider only the stabilisation of the complex. Dynamical aspects
concerning the encounter of the two proteins in the cell and the
formation of the complex are not incorporated. The generic model we use
is built on observations of (universal) features of rigid
protein-protein recognition so that the physics which different
recognition processes have in common is captured in the model.

We apply a coarse-grained point of view on the level of both the sequence of the
amino acids on the so-called recognition sites of biomolecules at the
mutual interface and the residue-residue interactions stabilising the
complex. The backbones of the proteins are assumed to undergo no
refolding during the association process. This is a justified
assumption for most protein-protein recognition processes, although
notable exceptions do exist
\cite{Kleanthous_2000,Peppas_2002,Janin_2007}. Motivated by the
observation that hydrophobicity is the major driving force in
molecular recognition
\cite{Kleanthous_2000,Jones_1996,Wodak_2003,Janin_2007} we describe
the type of the residue at the position $i = 1,\ldots,N$ of the
recognition site by a binary variable \cite{Behringer_etal_2006,
  Behringer_etal_2007b} where one of the two values represents a
hydrophobic residue and the other one a polar residue. Note, that an
eigenvalue decomposition of the Miyazawa-Jernigan matrix leads to an
approximate parameterisation of residue-residue interactions by an
Ising-like energy term with discrete variables that can take on two
distinct values \cite{Li_1997}. This gives additional justification to
the use of HP-models for the residue-residue interactions. Denoting
the type of the residue at position $i$ of the recognition site of one
of the two molecules by $\sigma_i \in \{ +1 (\mb{hydrophobic}), -1
(\mb{polar}) \}$ the residue sequence on the recognition site with $N$
residues is then specified by $\sigma = (\sigma_1, \ldots, \sigma_N)$.
Similarly the type of residue at position $i$ of the recognition site
of the interaction partner is specified by $\theta = (\theta_1,
\ldots,\theta_N)$ with $\theta_i \in \{\pm1 \}$.

We then model the energetics at the two-dimensional contact interface
of the two biomolecules by
\begin{equation}
\label{behringer:HP-coop}
\mathcal{H}_{\sts \mb{int}}(\sigma, \theta;S) = - \varepsilon\sum_{i=1}^{N}
\frac{1+S_i}{2} \sigma_i \theta_i - J
\sum_{\left<i,j\right>} S_i S_j
\end{equation}
where the energy contributions of the contact between two residues
across the interface are summed up.  The variable $S_i$ takes on the
two discrete values $\pm 1$ and describes the fit of the shape of the
molecules at position $i$ of the interface, for a poor fit, i.\,e.
$S_i = -1$, we assume no contribution to the stabilising energy. The
variable $S$ models the influence of a (local) rearrangement of the
amino acid side chains on a microscopic level when the complex is
formed \cite{Kleanthous_2000,Wodak_2003,Janin_2007}. Note that such
rearrangements are observed even if the tertiary structures of the
proteins remain unaltered upon complex formation.  Apart from the
direct contact energy with strength $\varepsilon$ the model
Hamiltonian (\ref{behringer:HP-coop}) contains an additional
cooperative interaction term where the quality of a residue-residue
contact couples to the structure in its neighbourhood. This term has
the effect that a locally good fit at some position in the interface
influences its neighbourhood \cite{Behringer_etal_2007b}.

In our idealised view of the interface each biomolecule contributes
with the same number $N$ of coarse-grained ``residues''. This
assumption is questionable for real interfaces, particularly for
curved interfaces different numbers of amino acids appear
\cite{Wodak_2003}. In Hamiltonian (\ref{behringer:HP-coop}) a residue
of one of the biomolecules interacts precisely with one residue on the
other molecule. This simplified assumption is also not valid for real
residues, in particular as different amino acids are of different
sizes so that a large residue can interact with several smaller amino
acids. However, one can think of a general partition of the interface
in $N$ contact patches of the same size on each of the biomolecules
where larger amino acids contribute to several patches whereas small
ones only to a few. A value of the hydrophobicity can then be
attributed to each of the patches on the biomolecules. Within such a
description the (free) energies can be approximated by the model
(\ref{behringer:HP-coop}).  For the sake of simplicity, however, we
stick to the expression ``residue'' in the following discussions.
We also note that solvation effects at the recognition
sites and the associated entropy changes are crucial when the complex
of two biomolecules is formed \cite{Gilson_1997,Jackson_2006}.  In the
adopted coarse-grained approach, however, it is assumed that all these
contributions are of comparable size for all proteins under
consideration.  Notice also that by reducing the interactions to the
hydrophobic effect solvation effects are already partially included in
HP-like models (on a formal level due to integrating out the solvent
degrees of freedom resulting in effective interaction constants like
$\varepsilon$ in (\ref{behringer:HP-coop})).

To study the recognition process between two rigid proteins we adopt a
two-stage approach. For a fixed target sequence
$\sigma^{(\sts\mb{t})}$ we first design an ensemble of probe molecules
$\theta$ at a design temperature $1/\beta_{\sts{\mb{D}}}$ in such a
way that the sequence $\theta$ should optimise the interface
energy. This design by equilibration leads to the distribution
$P(\theta | \sigma^{(\sts\mb{t})}) = \frac{1}{Z_{\sts \mb{\tiny D}}}
\sum_{S} \exp\left(-\beta_{\sts\mb{\tiny D}}
  \mathcal{H}(\sigma^{(\sts\mb{t})}, \theta;S) \right)$.  This first
step should mimic evolutionary processes or the design of artificial
molecules in biotechnological applications. The quality of the design
can be quantified by evaluating the average
$\left<K\right>_{P(\theta|\sigma^{(\sts\mb{t})})}$ of the overlap $K=
\sum_i \sigma^{(\sts\mb{t})}_i\theta_i$ of the sequence of the probe
molecules with the previously fixed target sequence. A large
$\left<K\right>_{P(\theta|\sigma^{(\sts\mb{t})})}$ then signals a high
complementarity of the two recognition sites in regard to the actual
recognition process of the two proteins. Notice that
$\left<K\right>_{P(\theta|\sigma^{(\sts\mb{t})})}$ is generally
dependent on the particular chosen target sequence
$\sigma^{(\sts\mb{t})}$.

In a second step the free energy difference of association at
temperature $1/\beta$ is calculated for the interaction of the probe
ensemble with the target molecule $\sigma^{(\sts\mb{t})}$ on the one
hand and a structurally different rival molecule
$\sigma^{(\sts\mb{r})}$ on the other hand. In this step the free
energy of the interaction
\begin{equation}
  F(\theta|\sigma^{(\alpha)}) = -\frac{1}{\beta} \ln \left( \sum_S \exp
  (-\beta\mathcal{H}(\sigma^{(\alpha)}, \theta;S)) \right)
\end{equation}
of the molecule $\sigma^{(\alpha)}$, $\alpha\in \{\mb{t }
(\mb{target}),\mb{r } (\mb{rival})\}$, with a particular probe
sequence $\theta$ has to be averaged with respect to the distribution
$P(\theta | \sigma^{(\sts\mb{t})})$ giving $F^{(\alpha)} =
\left<F(\theta|\sigma^{(\alpha)})\right>_{P(\theta|\sigma^{(\sts\mb{t})})}$.
This leads finally to the free energy difference $\Delta
F(\sigma^{(\sts\mb{t})},\sigma^{(\sts\mb{r})}) = F^{\sts
  \mb{(\sts\mb{t})}} - F^{\sts\mb{(\sts\mb{r})}}$.  In order to value
the recognition ability of the system the free energy difference
$\Delta F $ is then averaged over all possible target and rival
sequences on their respective recognition sites:
\begin{equation}
  \left [\Delta F\right]_{\sigma^{(\sts\mb{t})}, \sigma^{(\sts\mb{r})}} =
  \sum_{\sigma^{(\sts\mb{t})}, \sigma^{(\sts\mb{r})}} W^{(\sts\mb{t})}(\sigma^{(\sts\mb{t})}) W^{(\sts\mb{r})}(
  \sigma^{(\sts\mb{r})}) \Delta F
\end{equation}
where the $W^{(\alpha)}$ denote the distributions of the sequence of
the target and rival molecules, respectively.  A negative $\left
  [\Delta F\right]_{\sigma^{(\sts\mb{t})}, \sigma^{(\sts\mb{r})}}$
then signals an overall preferential interaction of the probe molecule
with the target leading to the desired selectivity of the recognition
process. In the following discussions square brackets indicate an
average over all possible target and rival sequences whereas pointed
brackets denote an average over the designed ensemble of probe
molecules.

Our approach can be roughly illustrated by the technologically
relevant case of developing a drug molecule with a high affinity to a
particular protein. The target molecule of our terminology corresponds
to a known protein which is responsible for a disease, for example,
with a well-located recognition site. Our design step then corresponds
to finding the most suitable drug molecule called probe in our
nomenclature. The subsequent testing step then models the
administration of the drug to an organism where additional proteins
(rival molecules) are present apart form the known protein the drug
molecule is supposed to bind to, so that all theses proteins can compete
for the drug molecules.

\section{Incorporating sequence correlations}
\label{kap:korrel}

In Hamiltonian (\ref{behringer:HP-coop}) only the energetics of the
contact interactions of residues across the interface between the two
interacting molecules is taken into account. However, the residues
that constitute the recognition sites on the proteins also interact
with each other, so that different sequences result in different
contributions to the total energy. Non-covalent hydrophobic-polar
contacts between neighbouring residues in the recognition sites, for
example, lead to unfavourable energy contributions. As a consequence
patches of several hydrophobic or polar residues are likely to show
up. Thus the probability of having a certain type of residue at
position $i$, say, in the recognition site depends on the type of the
residues in the neighbourhood of $i$, so that the sequences are
correlated. Indeed the appearance of patches of residues of a similar
hydrophobicity can be observed in the majority of protein-protein
interfaces \cite{Larsen_etal_1998}.

On a formal level, correlations can be incorporated by introducing,
apart from the contact energy $\mathcal{H}_{\sts \mb{int}}$ at the
interface, an additional correlation term $\mathcal{H}_{\sts
  \mb{cor}}$ to the Hamiltonian.  Note that in principle correlation
energies also show up in the interior of the proteins and in turn
induce correlations on the surface of the molecules. In this work,
however, we are only concerned with the interaction between two
proteins which depends on the nature of the residues that constitute
the recognition sites. We thus do not consider these further
distributions of interior (or other surface) residues explicitly.

Focusing on the sequence $\theta$ of the probe molecules for the
discussion we consider the following correlation energy:
\begin{equation}
\label{eq:correlhamilton}
  \mathcal{H}_{\sts \mb{cor}} = -\gamma_{\sts\mb{p}}
  \sum_{\left<i,j\right>} \theta_i\theta_j - \mu_{\sts \mb{p}} \sum_i \theta_i.
\end{equation}      
The first sum extends over all neighbouring residues in contact and
hence represents the interactions due to hydrophobicity so that the
associated parameter $\gamma_{\sts\mb{p}}$ thus controls the
corresponding (nearest-neighbour) correlations. These correlation
interactions lead to the formation of extended patches of either
hydrophobic or polar residues in the recognition sites. The
characteristic extensions of these patches can be interpreted as a
measure of the correlation length $\lambda_{\sts \mb{p}}$. In the
second contribution the hydrophobicity of the recognition site couples
to the parameter $\mu_{\sts\mb{p}}$ which therefore controls the
overall number of hydrophobic residues. The design step then gives the
probability of a certain probe sequence $\theta$ for a given target
sequence $\sigma^{(\sts\mb{t})}$. This probability distribution for
the probe ensemble is then generally given by
\begin{equation} 
P(\theta | \sigma^{(\sts\mb{t})}) = \frac{1}{\mathcal{N}} \exp (-\beta_{\sts\mb{D}}\mathcal{H}_{\sts \mb{int}} - \mathcal{H}_{\sts \mb{cor}})
\end{equation}
where $\mathcal{N}$ denotes the normalisation.  In general this
probability depends on the particular sequence $\sigma^{(\sts\mb{t})}$
of the recognition site of the given target. Note that the
contributions from the correlation energy are considered not to be
subjected to thermal fluctuations as only the rearrangement variable
$S$ is assumed to equilibrate.

After the average over the probe ensemble has been carried out the
free energy difference $\Delta F (\sigma^{(\sts\mb{t})},
\sigma^{(\sts\mb{r})})$ for a given target-rival pair depends on the
parameters $\gamma_{\sts\mb{p}}$ and $\mu_{\sts\mb{p}}$. For the final
average over the possible target and rival molecules sequences with
particular correlation properties are considered. Formally the
corresponding probability distributions for $\alpha \in \{\mb{t
  (target), r (rival)}\}$ are given by
\begin{equation}
\label{eq:verteiltarget}
  W^{(\alpha)}(\sigma^{(\alpha)}) \sim \exp (-\mathcal{H}_{\sts \mb{cor}}(\sigma^{(\alpha)}))
\end{equation}
with associated parameters $\gamma_{\alpha}$ for the (nearest-neighbour)
correlations and $\mu_{\alpha}$ for the overall hydrophobicity
\begin{equation}
  H_\alpha = N h_\alpha = \left[\sum_i \sigma_i^{(\alpha)}\right]_{W^{(\alpha)}}.
\end{equation}

For the investigation of the influence of sequence correlations on
molecular recognition in our model we adopted the following strategy.
For a fixed pair of target and rival sequences the probe ensemble will
be generated for the parameters $\gamma_{\sts\mb{p}}$ and
$\mu_{\sts\mb{p}}$ which in turn determine the correlation length
$\lambda_{\sts\mb{p}}$.  Note that the generated probe molecules are
not perfect with respect to the target molecule due to evolutionary
processes leading to defects. Then the recognition ability is assessed
by evaluating the free energy difference $\Delta F
(\sigma^{(\sts\mb{t})},\sigma^{(\sts\mb{r})})$ for the given
target-rival pair. This free energy difference is then averaged over
all possible target-rival pairs, where similarly to the probe molecule
the associated parameters $\gamma_\alpha$ and $\mu_\alpha$ determine
the correlation lengths $\lambda_\alpha$. By this approach the overall
recognition ability $\left [\Delta F\right]_{\sigma^{(\sts\mb{t})},
  \sigma^{(\sts\mb{r})}}
(\lambda_{\sts\mb{t}},\lambda_{\sts\mb{r}},\lambda_{\sts\mb{p}})$ is
hence computed as a function of the correlation lengths (and
hydrophobicities) of the target and rival molecules and of the
predesigned probe molecules. For given correlation lengths
$\lambda_{\sts\mb{t}}$ and $\lambda_{\sts\mb{r}}$ of the target and
rival molecules, respectively, the correlation length
$\lambda_{\sts\mb{p}}$ of the probe molecules is then varied to find
the optimum recognition ability.

\section{Uncooperative model}
\label{kap:unkoop}

The interaction energy (\ref{behringer:HP-coop}) at the interface
between the two proteins comprises apart form the direct contact
contributions due to hydrophobicity additional cooperative terms where
the rearrangements of neighbouring amino acid side chains couple to
each other. In this section we set the corresponding interaction
constant $J$ to zero and consider only the direct hydrophobic energy
contributions. The total Hamiltonian for the interface energy between
a molecule with the sequence $\sigma$ and the probe molecule $\theta$
thus reads
\begin{equation}
\label{eq:h-unkooperativ}
  \mathcal{H}_{\sts \mb{}} (\sigma, \theta;S) =  \mathcal{H}_{\sts
    \mb{int}}  + \frac{1}{\beta}\mathcal{H}_{\sts \mb{corr}} = - \varepsilon\sum_{i=1}^{N}
\frac{1+S_i}{2} \sigma_i \theta_i - \frac{\gamma_{\sts\mb{p}}}{\beta} \sum_{\left<i,j\right>}
  \theta_i\theta_j - \frac{\mu_{\sts\mb{p}}}{\beta} \sum_i\theta_i. 
\end{equation}
As the interaction variable $S_i$ at position $i$ does not couple to
the variables at other positions $j \neq i$ of the interface the
corresponding thermal average can be carried out resulting in an
effective Hamiltonian that depends only on the sequence variables any
more. Including the contributions from the correlation energies it is
given by
\begin{equation}
\label{eq:heffektiv}
  \mathcal{H}_{\sts \mb{eff}} (\sigma, \theta) =
  -\frac{\varepsilon}{2}
\sum_i \sigma_i\theta_i - \frac{\gamma_{\sts\mb{p}}}{\beta} \sum_{\left<i,j\right>}
  \theta_i\theta_j - \frac{\mu_{\sts\mb{p}}}{\beta} \sum_i\theta_i + \mb{const}.
\end{equation} 
Here we have used the fact that $\cosh (\beta\varepsilon
\sigma_i\theta_i) = \cosh(\beta\varepsilon)$ for all choices of
$\sigma_i$ and $\theta_i$. The constant in (\ref{eq:heffektiv}) is
temperature dependent, however, as we are only concerned with the
effect of correlations on the molecular recognition ability, we fix
the temperature and thus can omit the constant.  The free energy for
the interaction between the sequences $\sigma$ and $\theta$ is
$F(\theta|\sigma) = -\frac{\varepsilon}{2} \sum_i \sigma_i \theta_i +
\frac{1}{\beta} \mathcal{H}_{\sts\mb{cor}}(\theta)$ and can now be
averaged over the possible probe sequences that are distributed
according to the probability $P(\theta|\sigma^{(\sts\mb{t})}) \sim
\exp (-\beta_{\sts \mb{}}\mathcal{H}_{\sts
  \mb{eff}}(\sigma^{(\sts\mb{t})}, \theta))$.  Note that the design
might be carried out at a temperature $\beta_{\sts \mb{D}}$ which is
different from the temperature $\beta_{\sts \mb{}}$ at which the
selectivity is determined. However, we are not interested in the
effect of a temperature variation in this work and therefore choose
$\beta_{\sts \mb{D}} = \beta_{\sts\mb{}}$.  The correlation energy
$\mathcal{H}_{\sts\mb{cor}}$ does not explicitly depend on the
sequence $\sigma^{(\alpha)}$ and hence when computing the free energy
difference between the interaction of the target molecule with the
probe ensemble on the one hand and the interaction of the rival
molecule with the probe ensemble on the other hand these correlation
contributions cancel and one ends up with
\begin{equation}
\label{eq:deltaf}
  \Delta F(\sigma^{(\sts\mb{t})}, \sigma^{(\sts\mb{r})}) = -\frac{\varepsilon}{2} \sum_i
  (\sigma_i^{(\sts\mb{t})} - \sigma_i^{(\sts\mb{r})}) \left< \theta_i\right>_{P(\theta|\sigma^{(\sts\mb{t})})}.
\end{equation}    
The free energy difference is hence determined by the difference of
the complementarity of the probe ensemble with the target sequence on
the one hand and the complementarity of the probe ensemble with the
rival sequence on the other hand.  Note also that the free energy
difference exhibits a dependence on the correlation parameters
$\gamma_{\sts\mb{p}}$ and $\mu_{\sts\mb{p}}$ (which enter the
distribution $P$ and hence influence the average hydrophobicity at
position $i$ of the recognition site of the probe molecule) and thus
on the correlation length $\lambda_{\sts\mb{p}}$.

To assess the overall recognition ability the free energy difference
(\ref{eq:deltaf}) has to be averaged over all target and rival
sequences which are distributed with respect to
(\ref{eq:verteiltarget}) with correlation Hamiltonians of the form
(\ref{eq:correlhamilton}). As the target and the rival sequences are
independent of each other, the averaged free energy difference is
therefore given by
\begin{eqnarray}
\label{eq:unkoop_freieener0}
  \left [\Delta F\right]
&=&
  -\frac{\varepsilon}{2} \sum_i \left[ \sigma_i^{(\sts\mb{t})}\left<
  \theta_i\right>_{P(\theta|\sigma^{(\sts\mb{t})})}\right]_{W^{(\sts\mb{t})}} 
+  \frac{\varepsilon}{2} N [h_{\sts\mb{p}}]_{W^{(\sts\mb{t})}}
h_{\sts\mb{r}}  \\
\label{eq:unkoop_freieener}
&=& -\frac{\varepsilon}{2}\left[\left<K\right>_{P(\theta|\sigma^{(\sts\mb{t})})}\right]_{W^{(\sts\mb{t})}} +
  \frac{\varepsilon}{2} N [h_{\sts\mb{p}}]_{W^{(\sts\mb{t})}} h_{\sts\mb{r}} 
\end{eqnarray} 
in terms of the complementarity of the probe ensemble and
hydrophobicities $h_{\sts\mb{p}}$ and $h_{\sts\mb{r}}$ of the probe
and rival molecule, respectively. The second term originates from the
interaction of the probe molecules with the rival molecule. It is only
determined by the respective hydrophobicities of the molecules and is
independent of the structure elements related to the hydrophobic and
polar patches of the recognition sites. Note that the hydrophobicity
$h_{\sts\mb{p}}$ hinges on the sequence of the target molecule. The
first term stems from interactions of the probe molecule with the
target molecule. This term depends sensitively on an appropriate
matching of the structure elements on the recognition sites and is
hence directly influenced by correlation effects in the corresponding
distributions of the hydrophobicity.

In the following subsections we use two methods to carry out the
remaining averages in (\ref{eq:unkoop_freieener}), namely numerical
Monte Carlo techniques and a mean field approximation. Larsen et al.
reported that basically two types of interfaces appear in
protein-protein complexes \cite{Larsen_etal_1998}. In the minority of
complexes the interface has a hydrophobic core which consists of a
single large patch and which is surrounded by a rim of polar
interactions with residual accessibility by solvent molecules.  For
the majority of complexes, however, the interface is made up by a
mixture of small hydrophobic patches and polar interactions. We thus
focus in the following discussions only on the situation where the
correlation lengths of the target and rival molecule, respectively,
are relatively small compared to the extension of the interface.

\subsection{Numerical results}
\label{kap:mc}

The remaining averages in expression (\ref{eq:unkoop_freieener}) of
the free energy difference --- first over the probe ensemble with the
distribution $P(\theta|\sigma^{(\sts\mb{t})})$ and then over the
target sequences with the distribution $W^{(\sts\mb{t})}$ --- can be
carried out numerically by means of Monte Carlo methods. For a given
target and rival sequence the quantities of interest (averaged
complementarity and free energy difference as a measure for
selectivity) are computed first. Then the final average over the
target sequences with fixed parameters $\gamma_{\sts\mb{t}}$ and
$\mu_{\sts\mb{t}}$ (and hence fixed correlation length
$\lambda_{\sts\mb{t}}$ and hydrophobicity $h_{\sts\mb{t}}$) is
evaluated. As we are interested in the recognition ability of the
system if the rival molecule is structurally very similar to the
target molecule, the same correlation parameters are used for the
average over the rival sequences and thus one has in particular
$h_{\sts\mb{r}}=h_{\sts\mb{t}}$.

The probe molecules are designed for different correlation parameters
$\gamma_{\sts\mb{p}}$. The probe sequence is optimised with respect to
the target sequence, thus we do not further restrict the
hydrophobicity and therefore set $\mu_{\sts\mb{p}}=0$. The correlation
parameter $\gamma_{\sts\mb{p}}$ can therefore be directly converted
into the correlation length $\lambda_{\sts\mb{p}}$. The (pseudo-)
correlation length for recognition sites of a finite extension is
computed to be the average size of clusters that are made up of
neighbouring residues of the same type.  In the following figures the
shown correlation length $\lambda_{\sts\mb{p}}$ is normalised in such
a way that its maximum possible value is one for a system where the
whole recognition site is made up of precisely one cluster with either
hydrophobic or polar residues.

Alternatively the correlation length of a finite system can be defined
by the second moment of an (appropriately normalised) correlation
function \cite{Binder_etal_1985}. However, both definitions lead to
the same qualitative behaviour of the correlation length as a function
of the varying correlation parameters. The correlation length
increases monotonically as a function of the correlation parameter
$\gamma_{\sts\mb{p}}$ and saturates for sufficiently large values.
Note also that in \cite{Jayaraman_2005} the correlations on a finite
surface where measured by a so-called patchiness which was defined to
be basically the (suitably normalised) expectation value of the
correlation energy $\sum_{\left<ij\right>}\theta_i\theta_j$ in terms
of our notation and convention.

For simplicity the systems considered for the Monte Carlo simulations
are of regular rectangular geometry and contain between 64 and 256
spin variables. Note that real recognition sites contain typically
30-40 residues, however, up to minor finite-size effects we find the
same qualitative behaviour for systems of different sizes. As
indicated in the introduction the energy contribution $\varepsilon$ of
a non-covalent bond is only slightly stronger than the thermal energy
at physiological conditions. We therefore typically choose
$\beta\varepsilon \geq\mathcal{O}(1)$. In the following results we
discuss the system with $\beta\varepsilon = 1$ if not stated
otherwise.

\begin{figure}[h!]
\begin{center}
\includegraphics[scale=0.285,angle=0]{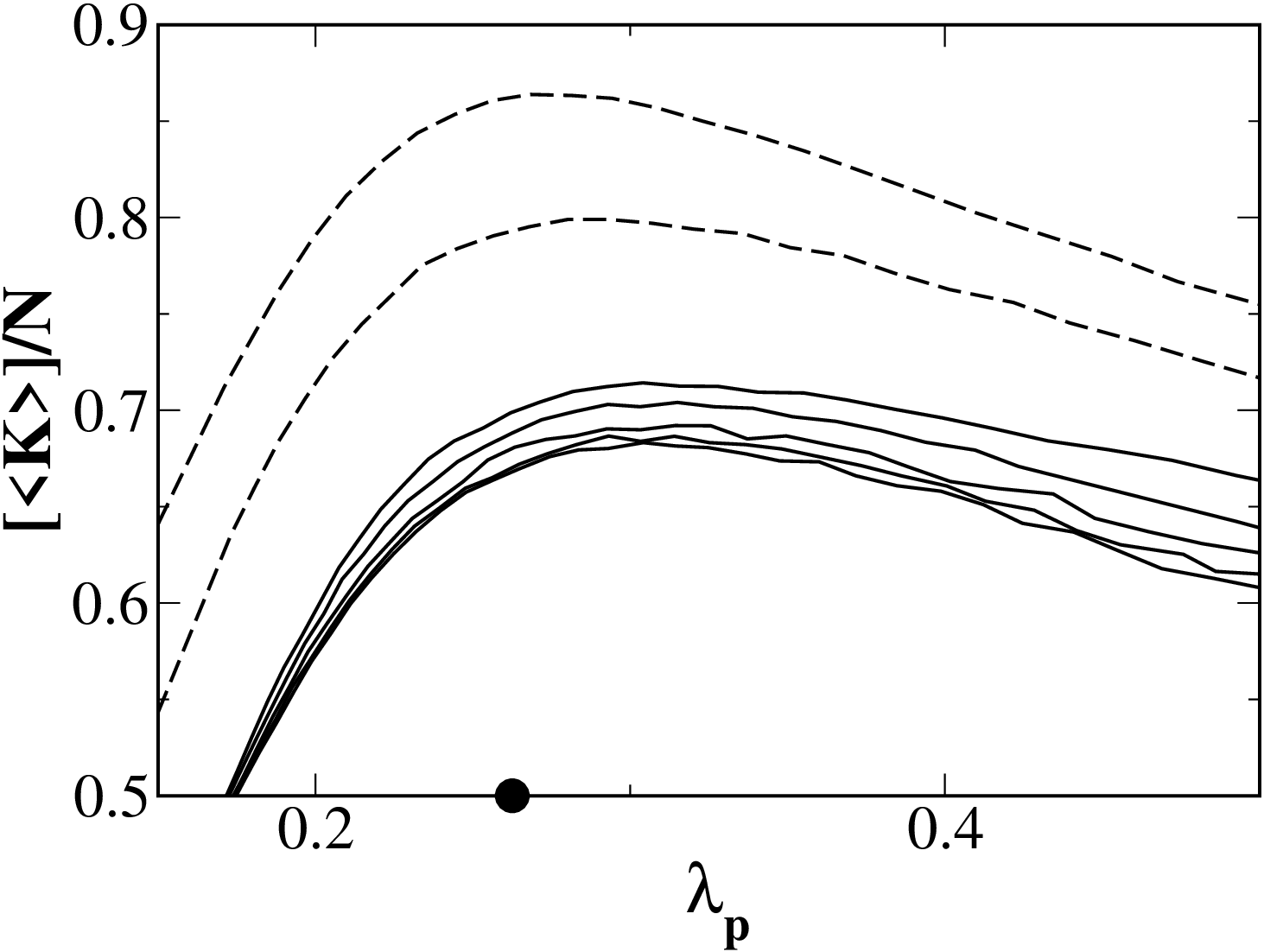}
\caption{\label{bild:comp_korrel} Average complementarity of the probe
  ensemble with $\beta_{\tiny\mb{}}\varepsilon = 1$ as a function of
  the correlation length for targets with different hydrophobicities
  (solid lines, from top to bottom, $h_{\sts \mb{t}}=0.5$, $0.4$,
  $0.3$, $0.2$ and $0.1$, the curve for $h_{\sts \mb{t}}=0.0$ is not
  shown as it is hardly distinguishable from the one with $h_{\sts
    \mb{t}}=0.1$ in the displayed range of
  $\lambda_{\sts\mb{p}}$). The correlation length of the targets is
  fixed to the value indicated by the black circle
  ($\lambda_{\sts\mb{t}}= 0.263$, corresponding e.\,g. to
  $\gamma_{\sts\mb{t}}=0.4$ for $h_{\sts \mb{t}}=0.0$). The optimum of
  the complementarity is slightly shifted to larger correlation
  lengths on the probe molecule. For the dashed curves
  $\beta_{\tiny\mb{}}\varepsilon = 1.5$ and $2.0$ (from the bottom
  up), again with $h_{\sts \mb{t}}=0.5$.}
\end{center}
\end{figure}

Consider a system with targets and rivals whose correlation length is
relatively small so that the recognition sites consist of a relatively
large number of rather small hydrophobic and polar patches.  We
investigated systems with hydrophobicities ranging from $h_{\sts
  \mb{t/r}} = 0.0$ to $h_{\sts \mb{t/r}} = 0.5$ and correlation
lengths between $\lambda_{\sts \mb{t/r}} = 0.2$ and $\lambda_{\sts
  \mb{t/r}} = 0.35$ (note that the uncorrelated system with
$\gamma_{\sts \mb{t/r}} = 0.0$ corresponds to a correlation length
larger than the minimum length $\lambda_{\sts \mb{t/r}} = 1/L$ for a
system with linear extension $L $ due to finite size effects). For all
the systems we find the same qualitative behaviour, we therefore
discuss exemplarily the system with $L =16$ and $\lambda_{\sts
  \mb{t/r}} = 0.263$ in the following.

\begin{figure}[h!]
\begin{center}
\includegraphics[scale=0.285,angle=0]{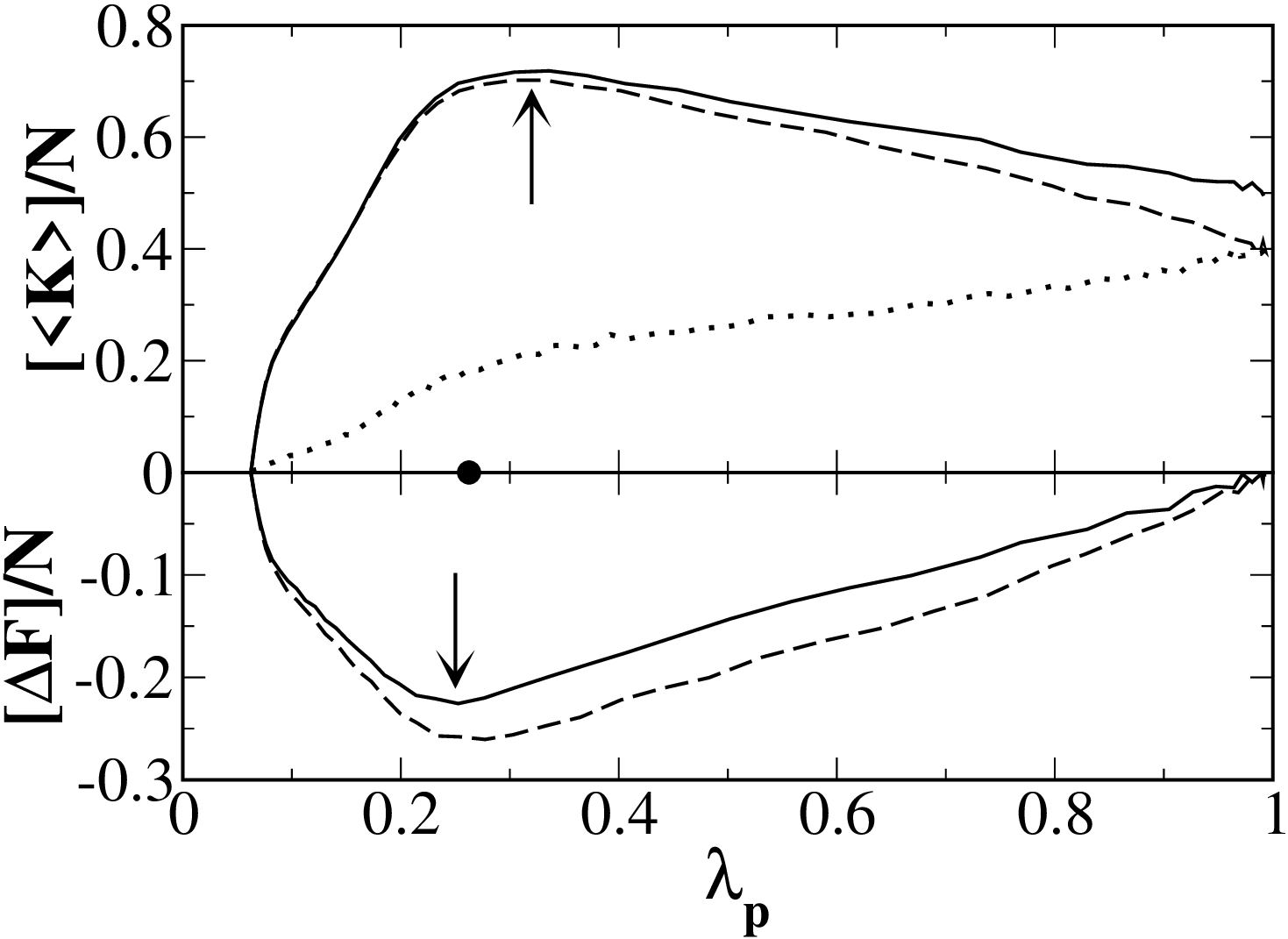}
\caption{\label{bild:freie_korrel} The complementarity $\left [\left<K
    \right>\right]/N$ and the free energy difference $\left[ \Delta
    F\right]/N$ as a function of the correlation length of the probe
  molecules. The correlation lengths of the target and rival molecules
  are fixed to the value shown by the circle
  ($\lambda_{\sts\mb{t}}=\lambda_{\sts\mb{r}} = 0.263$), the
  corresponding hydrophobicities are
  $h_{\sts\mb{t}}=h_{\sts\mb{r}}=0.5$ (solid line) and 0.4 (dashed
  line).  Compared to the optimum for the design of the probe
  molecules, the optimum of the recognition ability is clearly shifted
  to smaller values of the correlation length on the probe molecule
  (optima indicated by arrows for $h_{\sts\mb{t}}=0.5$). Additionally,
  the complementarity of the probe ensemble with respect to the rival
  molecules is shown for $h_{\sts\mb{t}}=0.4$ (dotted line). Notice
  that the system for the shown data has a linear extension $L=16$ and
  hence the minimum possible correlation length is
  $\lambda_{\sts\mb{p}} \approx 0.06$, the uncorrelated system with
  $\gamma_{\sts\mb{p}} = 0$ has $\lambda_{\sts\mb{p}} \approx 0.16$.}
\end{center}
\end{figure}

In figure \ref{bild:comp_korrel} the average complementarity of the
designed probe molecules is shown as a function of varying correlation
length $\lambda_{\sts\mb{p}}$ of the recognition site of the probe
molecules for different hydrophobicities of the target molecules. It
has to be noted first, that the complementarity (as well as the
selectivity, which is discussed below) is first enhanced by increasing
correlations, reaches an optimum and finally decreases again. The
probe molecules are expected to have a maximum complementarity if the
patches of hydrophobic and polar residues on the target are matched by
corresponding patches on the probe. However, the optimisation of the
probe ensemble is carried out at a finite temperature and therefore
thermal fluctuations limit the complementarity due to defects in the
distribution of the interaction partner as the patches fray out at
their boundaries. The position of the maximum of the average
complementarity, that corresponds to the optimum choice of the
correlation length of the probe molecules, is shifted to slightly
larger values compared to the fixed correlation length of the target
molecule. This signals the fact that a slightly larger correlation
length compensates the appearance of defects in the boundaries of the
patches during the design step and thus increases the
complementarity. This effect is less pronounced if the temperature is
decreased as defects appear more seldom.  Notice also that the average
complementarity tends to the fixed hydrophobicity $h_{\sts\mb{t}}$ of
the target in the limit $\lambda_{\sts\mb{p}} \to 1$ as in this case
the recognition site of the probe is made up of hydrophobic residues
only (compare figure \ref{bild:freie_korrel}).

\begin{figure}[h!]
\begin{center}
\includegraphics[scale=0.285,angle=0]{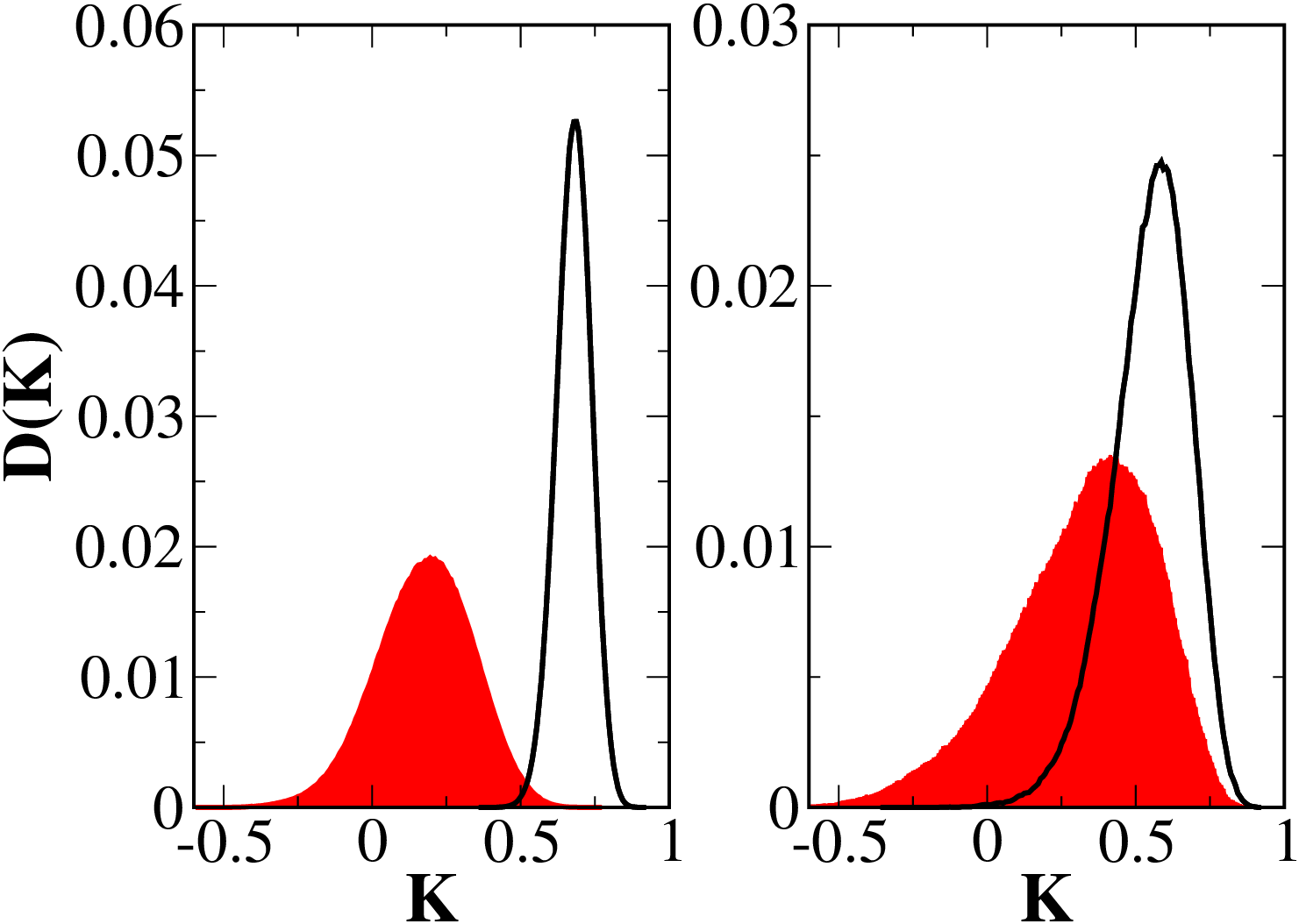}
\caption{\label{bild:kompdist} Distribution of the complementarity of
  the probe ensemble with respect to the target molecules (solid line)
  and the rival molecules (shaded curve) for different correlation
  lengths on the recognition site of the probe molecules. On the left
  hand side the correlation length $\lambda_{\sts\mb{p}} = 0.25$, on
  the right hand side $\lambda_{\sts\mb{p}} = 0.75$. The
  hydrophobicities of the target and rival molecules are
  $h_{\sts\mb{t}}=h_{\sts\mb{r}}=0.4$, the correlation lengths are
  $\lambda_{\sts\mb{t}}=\lambda_{\sts\mb{r}} = 0.263$ in each case.}
\end{center}
\end{figure}

For the uncooperative model (\ref{behringer:HP-coop}) of the direct
contact energy at the interface between the biomolecules the free
energy difference is determined by the difference in the
complementarity of the probe ensemble with respect to the target
molecules and the rival molecules, respectively (compare relation
(\ref{eq:deltaf})). In figure \ref{bild:freie_korrel} (upper part) the
complementarity with the rival molecules is shown in comparison with
the one with respect to the target as a function of the correlation
length $\lambda_{\sts\mb{p}}$. The probe ensemble is always more
complementary to the target, with respect to which it has been
optimised during the design step. For an increasing correlation length
on the probe molecule the complementarity with respect to the rival
sequence is increased until it finally reaches the maximum possible
value for $\lambda_{\sts\mb{p}} \to 1$. In this case the probe is not
structured any more and hence cannot discriminate between different
sequences any more. In figure \ref{bild:kompdist} the distribution
$D(K)$ of the complementarity parameter with respect to the target and
with respect to the rival molecules (averaged over all target and
rival sequences) are compared for two different correlation
lengths. For probe molecules with small structure elements with a
characteristic length in the proximity of the optimum value the two
distributions are well separated and hence the probe can discriminate
the two molecules. For increasing correlation length and hence
diminishing structuring of the probe molecules the two distributions
approach each other and therefore selectivity is decreased. This comes
along with a broadening of the distributions when going away from
correlation lengths that correspond to the optimum conditions for the
selectivity. For $\lambda_{\sts\mb{p}} \to 1$ to two distributions
become eventually identical. Similarly, the two distributions are
converging towards each other when the correlation length is decreased
to the minimum possible value.

Figure \ref{bild:freie_korrel} shows the free energy difference of the
interaction of the probe molecules in a system with target and rival
molecules, again as a function of the correlation length of the probe
molecules. Note that the hydrophobicity $h_{\sts\mb{p}}$ in
(\ref{eq:unkoop_freieener}) exhibits a dependence on
$\lambda_{\sts\mb{p}}$.  For $\lambda_{\sts\mb{p}} \to 1$ the free
energy difference has to vanish as the probe molecule consists only of
amino acids of the same class in this case and hence it can not
distinguish on average between different sequences any more. The
minimum of the free energy difference corresponds to a system with
optimum recognition ability.  The numerical results show that for
recognition sites of the target with an excess of hydrophobic residues
the optimum of the recognition ability is clearly shifted to smaller
values of the correlation length compared to the appearance of the
optimum in the design of the probe molecules. The reason for this
shift lies in the fact that the structure elements of the recognition
sites influence the contributions of the target-probe interactions to
the free energy difference whereas the rival-probe interactions do not
feel these structure elements. A smaller correlation length implies
the appearance of an increased number of smaller patches on the
recognition site of the probe molecule and hence an entropic benefit
for the interaction with the target due to more possible ways to align
each other favourably.  This effect does not contribute to the free
energy for the rival-probe interactions as it is insensitive to a
matching of structure elements (compare relation
(\ref{eq:unkoop_freieener}) and the discussion there). The emergence
of the shift of the optimum correlation length also means that the
design of the probe molecules has not to be carried out as effectively
as one might expect naively.  Therefore the system is at liberty to
carry out the design not at the possible optimum way without losing
the optimum recognition ability.

\begin{figure}[h!]
\begin{center}
\includegraphics[scale=0.285,angle=0]{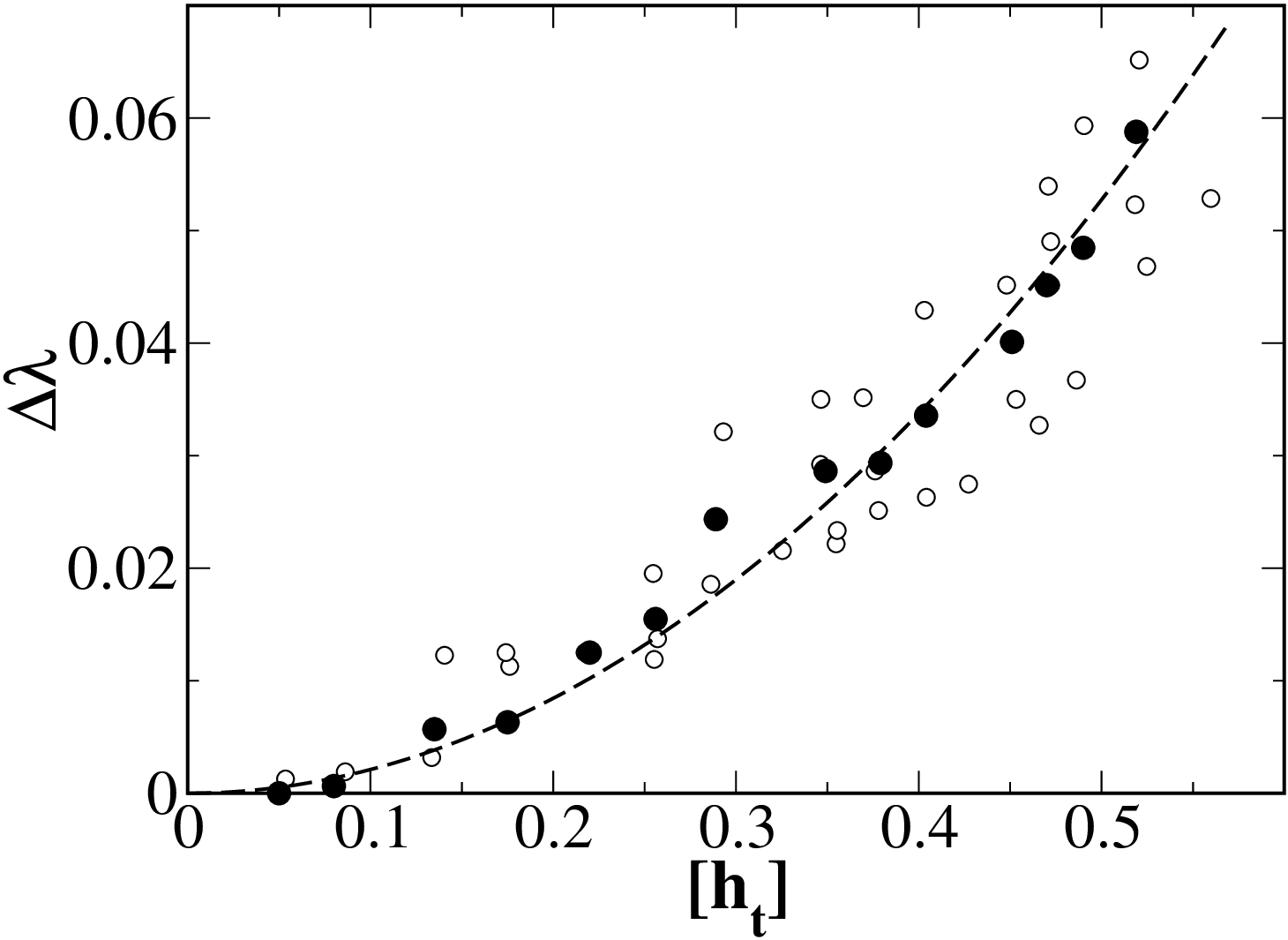}
\caption{\label{bild:deltafreie_korrel} The shift of the optimum value
  of the correlation length for the recognition ability compared to
  the optimum value for the complementarity as a function of the
  hydrophobicity of the target (note that
  $h_{\sts\mb{t}}=h_{\sts\mb{r}}$). Instead of error bars some of the
  results from the Monte Carlo runs (open circles) are shown together
  with the results of the analysis of the data (full circles). The
  dashed curve is a quadratic fit to the data (see discussion in
  section \ref{kap:meanfield}).}
\end{center}
\end{figure}

Interestingly this shift of the optimum correlation length depends on
the value of the hydrophobicity of the target and rival
molecule. Figure \ref{bild:deltafreie_korrel} shows that the shift
vanishes for recognition sites with the same number of hydrophobic and
polar residues (as is clear form relation (\ref{eq:unkoop_freieener}))
and increases with increasing hydrophobicity. Note that in nature
recognition sites with different hydrophobicities show up for proteins
with different biological function. In enzyme-inhibitor complexes one
typically finds largely hydrophobic interfaces whereas the
hydrophobicity in antibody-antigen interfaces is significantly lowered
\cite{Wodak_2003,Janin_2007}.

\begin{figure}[h!]
\begin{center}
\includegraphics[scale=0.285,angle=0]{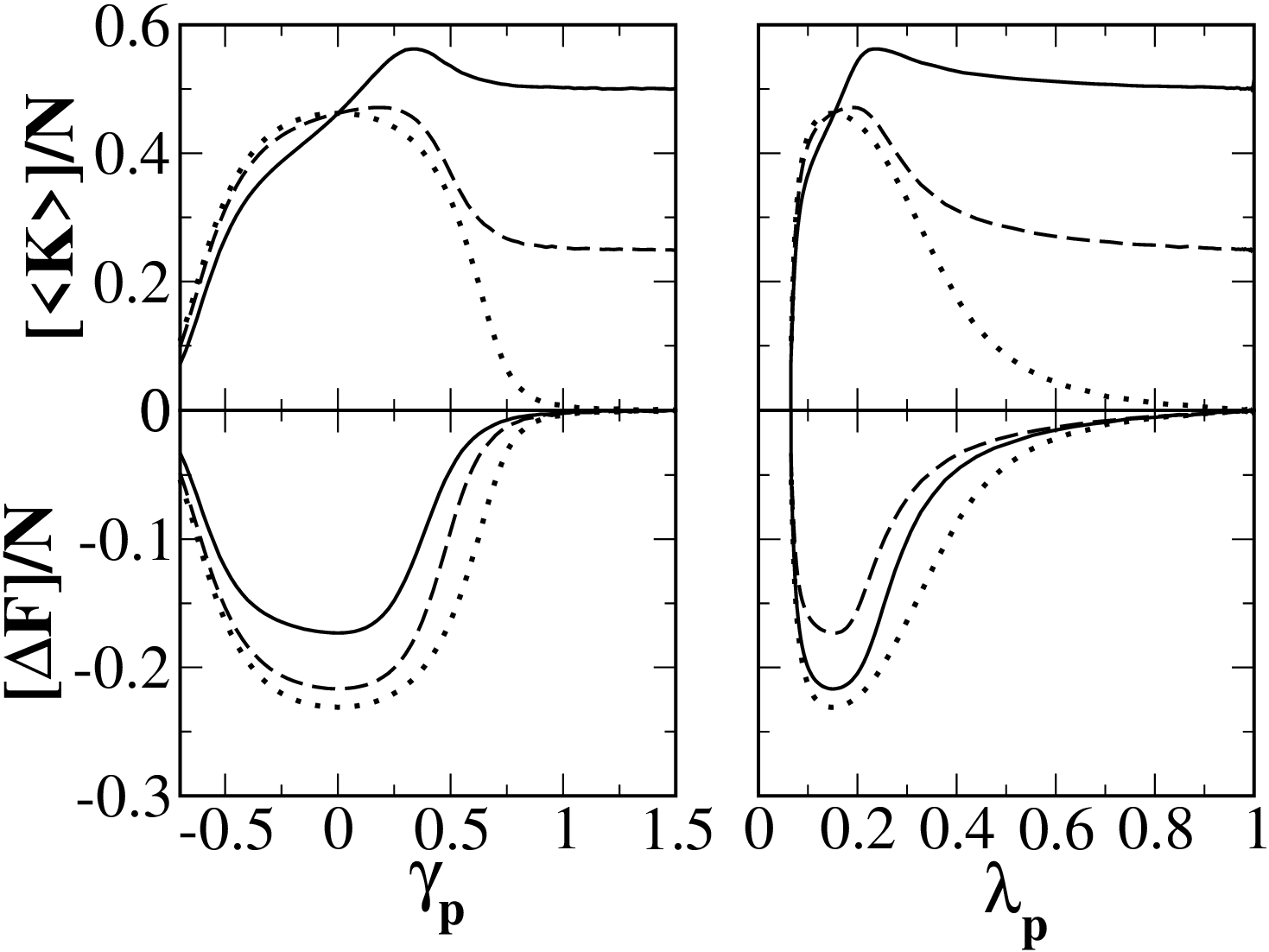}
\caption{\label{bild:freie_unkorrel} The complementarity $\left
    [\left<K \right>\right]/N$ and the free energy difference $\left[
    \Delta F\right]/N$ as a function of the correlation parameter
  $\gamma_{\sts\mb{p}}$ and of the correlation length
  $\lambda_{\sts\mb{p}}$, respectively, of the probe molecules. The
  correlation parameters of the target and rival molecules are set to
  zero, the corresponding hydrophobicities are fixed to the values
  $h_{\sts\mb{r}}=h_{\sts\mb{t}}=0.5$ (solid curve), 0.25 (dashed
  line) and 0.0 (dotted line). The free energy difference has an
  optimum for the correlation parameter $\gamma_{\sts\mb{p}} = 0.0$,
  the optimum complementarity, however, is shifted to larger values.}
\end{center}
\end{figure}

Although the recognition sites in real systems show always extended
patches of either hydrophobic or polar amino acids
\cite{Larsen_etal_1998} we briefly discuss systems where no nearest
neighbour correlations appear in the distribution of the residues on
the target and rival molecule. As a consequence the recognition site
is rather diffuse on average concerning the distribution of hydrophobic
and polar residues. The hydrophobicity of the corresponding
recognition sites is nevertheless fixed to a certain value and the
correlation length due to nearest neighbour correlations is varied on
the recognition site of the probe molecules to find the optimum
selectivity. The results for different hydrophobicities are depicted
in figure \ref{bild:freie_unkorrel}. The correlation parameter at
which the optimum complementarity of the probe molecules with respect
to the target molecules shows up depends on the hydrophobicity of the
target and is shifted to values larger than zero for positive
hydrophobicities. In this case the probe molecules prefer a
correlated, i.\,e. patch-structured surface although the target
surface is uncorrelated and thus unstructured.  The free energy, on the other hand, has
always its optimum for uncorrelated probe molecules. So again the
design need not be carried out in the optimal way, but
correlations will not enhance selectivity as in the case of correlated
targets and rivals.

Finally we compare our results to the findings of the work by Lukatsky
and Shakhnovich who investigated the influence of correlated density
distributions at the interface between biomolecules
\cite{Lukatsky_2008}. From their study they deduced that the presence
of correlations is a basic principle for recognition between proteins
and lead to an enhanced probability to find such interfaces as hub-hub
interactions in protein-protein networks. In our work we consider
correlations in the distribution of hydrophobic and polar residues
within the surface of the biomolecules. We basically reach the same
conclusions as Lukatsky and Shakhnovich. The corresponding
correlations lead to lower binding energies for moderately correlated
interfaces as is indicated by the increase of the averaged
complementarity as shown in figures \ref{bild:comp_korrel} and
\ref{bild:freie_korrel}. This points to a universal importance of
(different) correlation effects to ensure the necessary specificity of
recognition processes. Our approach contains an additional design step
where the two recognising proteins are optimised with respect to each
other. Note that the expression ``design'' has been used in
\cite{Lukatsky_2008} to refer to the emergence of correlations.

\subsection{Mean field approximation}
\label{kap:meanfield}

The averages in expression (\ref{eq:unkoop_freieener}) of the free
energy difference can not be evaluated analytically, however, progress
can be made by applying a mean field approximation. Introducing the
variable $k_i = \frac{\mu_{\sts\mb{p}}}{\beta}
+\frac{\varepsilon}{2}\sigma^{(\sts\mb{t})}_i$ the effective
Hamiltonian that describes the distribution of the sequence of the
probe molecules after the design step has been carried out is given by
\begin{equation}
\mathcal{H}_{\sts \mb{eff}} (\sigma, \theta) =
   - \frac{\gamma_{\sts\mb{p}}}{\beta} \sum_{\left<i,j\right>}
  \theta_i\theta_j -  \sum_i k_i\theta_i 
\end{equation}
dropping an irrelevant temperature-dependent constant. The variable
$k_i$ can be interpreted as a random variable whose probability is
determined by the distribution $W^{(\sts\mb{t})}$ of the target
sequence. The system can therefore be viewed as a random field Ising
model. The mean field treatment in the form of the equivalent
neighbour approximation amounts to replacing $\mathcal{H}_{\sts
  \mb{eff}}$ by
\begin{equation}
\mathcal{H}_{\sts \mb{eff}}^{(\sts \mb{MF})} (\sigma, \theta) =
   - \frac{\gamma_{\sts\mb{p}}}{2N\beta} \left(\sum_i
  \theta_i\right)^2 -  \sum_i k_i\theta_i.
\end{equation}
The expectation value $\left<
  \theta_i\right>_{P(\theta|\sigma^{(\sts\mb{t})})}$ in
(\ref{eq:unkoop_freieener0}) is then given by the derivative 
\begin{equation}
 \left<
  \theta_i\right>_{P(\theta|\sigma^{(\sts\mb{t})})} = -\frac{1}{N}\frac{\partial}{\partial
  k_i} G_{\sts \mb{eff}}
\end{equation}
where the effective free energy $G_{\sts \mb{eff}}$ is related to the
Hamiltonian $\mathcal{H}_{\sts \mb{eff}}^{(\sts \mb{MF})}$ by
\begin{equation}
  G_{\sts \mb{eff}} = -\frac{1}{\beta_{\sts{\mb{}}}}\ln Z_{\sts
  \mb{eff}} 
\end{equation}
with $Z_{\sts \mb{eff}} = \sum_\theta
\exp(-\beta_{\sts{\mb{}}}\mathcal{H}_{\sts \mb{eff}}^{(\sts
  \mb{MF})})$.  The effective partition function $Z_{\sts \mb{eff}}$
can be calculated in the large $N$ limit by first using the identity
\begin{equation}
\label{eq:hubbard}
   \exp\left(\frac{a}{2N} x^2\right) = \int\limits_{-\infty}^{+\infty}
   \mb{d} y \sqrt{\frac{Na}{2 \pi}}                                                               
  \exp\left(-\frac{Na}{2}y^2 + axy\right),     
\end{equation}
(with $a>0$) so that the variable $x := \sum_i\theta_i$ appearing
quadratically in the Boltzmann factor of $Z_{\sts \mb{eff}}$ is
linearised and hence the summation over $\theta$ can by carried
out. The price to pay for this linearisation is the introduction of
the auxiliary variable $y$.  Omitting irrelevant prefactors the
effective partition function is then given by
\begin{equation}
 \label{eq:Z-integral}
  Z_{\sts \mb{eff}}\sim 
  \int\limits_{-\infty}^{+\infty} \mb{d} y                                                           
  \exp\left( N\mathcal{A}(y,k) \right) 
\end{equation}
with the argument
\begin{equation}
  \mathcal{A}(y,k) = -\frac{\gamma_{\sts\mb{p}}}{2}y^2 +\frac{1}{N} \sum_i
  \ln\cosh\left( \gamma_{\sts\mb{p}} y + \beta k_i\right)
\end{equation}
where $k$ denotes the configuration $(k_1, \ldots,k_N)$.  The Laplace
method allows an asymptotic evaluation of (\ref{eq:Z-integral}) in the
large $N$ limit leading to
\begin{equation}
  G_{\sts \mb{eff}} = N\mathcal{A}(y_0,k) = -N\frac{\gamma_{\sts\mb{p}}}{2}y_0^2 +\frac{}{} \sum_i
  \ln\cosh\left( \gamma_{\sts\mb{p}} y_0 + \beta k_i\right)
\end{equation}
with the so-called mean field $y_0$ determined by the saddle point
equation
\begin{equation}
\label{eq:sattel}
  y_0 = \frac{1}{N} \sum_i\tanh \left( \gamma_{\sts\mb{p}} y_0 + \beta
  k_i\right).
\end{equation}
Note that the mean field depends explicitly on the sequence
$\sigma^{(\sts\mb{t})}$ of the recognition site of the target.
Having computed an expression for the effective free energy $G_{\sts
  \mb{eff}}$ one can now calculate the desired average 
\begin{equation}
 \left<
  \theta_i\right>_{P(\theta|\sigma^{(\sts\mb{t})})} = -\frac{1}{N}\frac{\partial}{\partial
  k_i} G_{\sts \mb{eff}} = \tanh\left( \gamma_{\sts \mb{p}}y_0 + \mu_{\sts
  \mb{p}} + \frac{\beta\varepsilon}{2}\sigma_i^{(\sts\mb{t})}\right).  
\end{equation}
Additionally one has $\sum_i \left<
  \theta_i\right>_{P(\theta|\sigma^{(\sts\mb{t})})} = Ny_0 $ so that the mean
field gives the expectation value of the hydrophobicity of the probe
ensemble. The free energy difference (\ref{eq:unkoop_freieener}) is then generally given by
\begin{eqnarray}
\label{eq:deltafvormit}
  \Delta F &=& -\frac{\varepsilon}{2}\left[\sum_i \sigma_i^{(\sts\mb{t})} \tanh\left( \gamma_{\sts \mb{p}}y_0 + \mu_{\sts
  \mb{p}}
+
  \frac{\beta\varepsilon}{2}\sigma_i^{(\sts\mb{t})}\right)\right]_{W^{(\sts\mb{t})}} 
+ \frac{\varepsilon}{2}Nh_{\sts\mb{r}}\left[y_0\right]_{W^{(\sts\mb{t})}}
\end{eqnarray}
where averages over the target and the rival sequences still have to
be carried out. 

Starting from expression (\ref{eq:deltafvormit}) these averages can be
carried out numerically. The mean field $y_0$, that is determined by
the saddle point equation (\ref{eq:sattel}), explicitly depends on the
target sequence $\sigma^{(\sts\mb{t})}$ and hence one has of the order
of $\mb{e}^N$ saddle point equations for a system with $N$ residues. A
particular configuration $\sigma^{(\sts\mb{t})}$, however, contains
$\Sigma^{(+)}$ hydrophobic residues and $\Sigma^{(-)}$ polar ones. For
such a configuration the saddle point equation is given implicitly by
the equation
\begin{eqnarray}
\label{eq:sattelmakro}
y_0(\Sigma^{(+)}, \Sigma^{(-)})  &=& \frac{\Sigma^{(+)}}{N} \tanh\left( \gamma_{\sts\mb{p}} y_0 + \mu_{\sts\mb{p}}
   +\frac{\beta\varepsilon}{2}\right) 
+  \frac{\Sigma^{(-)}}{N} \tanh\left(
 \gamma_{\sts\mb{p}} y_0 + \mu_{\sts\mb{p}} -\frac{\beta\varepsilon}{2}\right).
\end{eqnarray}
and hence the mean field depends only on the numbers $(\Sigma^{(+)},
\Sigma^{(-)})$ for a given configuration. This observation drastically
reduces the number of saddle point equations. The remaining equations
can be solved using computer algebra programmes, the average with
respect to the distribution $W^{(\sts\mb{t})}$ can be carried out
afterwards. A distribution $W^{(\sts\mb{t})}$ of the form
(\ref{eq:verteiltarget}) can be expressed in terms of the density of
states $\Omega(\Sigma^{(+)}, \Sigma^{(-)}, E)$ specifying the number
of target configurations that are compatible with the macroscopic
parameters $\Sigma^{(+)}, \Sigma^{(-)}$ and $E$, where $E$ denotes the
correlation energy. For fairly small systems this density of states
can be calculated exactly by suitable enumeration algorithms
\cite{Binder_1972}, for large systems effective Monte Carlos
techniques can be applied \cite{Wang_2001,Hueller_2002,Landau_2004}.

The mean field treatment reproduces the qualitative results of the
numerical investigations discussed in subsection \ref{kap:mc}. For
instance, the complementarity of the probe ensemble and the free
energy difference as a measure of the recognition ability of the
probe-target system in the presence of a rival molecule can now be
worked out as a function of the correlation parameter
$\gamma_{\sts\mb{p}}$. Again a characteristic shift of the optimum
correlation parameter and hence correlation length for the two
quantities can be observed in accordance with the above discussed
numerical Monte Carlo findings.

The mean field result can be used to consider the case of a small
correlation parameter $\gamma_{\sts\mb{p}}$ (with
$\mu_{\sts\mb{p}}=0$) in more details. The implicit saddle point
equation (\ref{eq:sattelmakro}) can be expanded into a power series in
$\gamma_{\sts\mb{p}}$ and solved up to oder $\gamma_{\sts\mb{p}}^2$.
This gives
\begin{equation}
\label{eq:sattelasymp}
 y_0 = h_{\sts\mb{t}}A + \gamma_{\sts\mb{p}}h_{\sts\mb{t}} B +
 \gamma_{\sts\mb{p}}^2 C_1 
\end{equation} 
with the numerical constants being $A = \tanh(\beta\varepsilon/2)$, $B
= \tanh(\beta\varepsilon/2)\mb{sech}^2(\beta\varepsilon/2)$ and $C_1 =
B (h_{\sts\mb{t}} - h^3_{\sts\mb{t}}\sinh^2(\beta\varepsilon/2))$.
Note that $y_0$ still depends on the particular sequence
$\sigma^{(\sts\mb{t})}$ of the target through the dependency on the
hydrophobicity $h_{\sts\mb{t}} = h_{\sts\mb{t}}
(\sigma^{(\sts\mb{t})}) = 1/N\sum_i\sigma^{(\sts\mb{t})}_i =
(2\Sigma^{(+)} - N)/N$.  Using (\ref{eq:sattelasymp}) the
complementarity of the probe ensemble averaged over all possible
target sequences can be computed up to order $\gamma_{\sts\mb{p}}^2$
giving
\begin{equation}
  \frac{1}{N}\left[
  \left<K\right>_{P(\theta|\sigma^{(\sts\mb{t})})}\right]_{W^{(\sts\mb{t})}} = A +
  \gamma_{\sts\mb{p}}[h_{\sts\mb{t}}^2] B +
  \gamma^2_{\sts\mb{p}}[h_{\sts\mb{t}}^2] C_2 
\end{equation} 
with $C_2 = B (1 - \sinh^2(\beta\varepsilon/2)) $. The complementarity
is determined in this limit by the second moment of the hydrophobicity
distribution of the target molecules and hence directly feels the
structure of the hydrophobic and polar patches on the recognition site
of the target. For sufficiently large $\beta\varepsilon$ this
expression has a maximum at a correlation parameter $\gamma_K =
-B/(2C_2)$.  Note that the position of the maximum is independent of
the properties of the distribution $W^{(\sts\mb{t})}$ of the target
sequences in the considered situation of a small correlation parameter
for the probe molecules, in particular it is independent of the chosen
hydrophobicity of the target molecules. The numerical Monte Carlo data
shown in figure \ref{bild:comp_korrel} seem to be in accordance with
this observation --- the data is shown as a function of the
correlation length, the maximum shows up at a fairly small correlation
length and hence a small correlation parameter.  The position where
the maximum appears is shifted to smaller values of the correlation
parameter and thus correlation length for increased
$\beta\varepsilon$. This is again confirmed by the numerical data in
figure \ref{bild:comp_korrel}. Similarly the free energy difference
can be work out as a second order Taylor polynomial in
$\gamma_{\sts\mb{p}}$.  It shows a minimum at a correlation parameter
$\gamma_F$. The shift $\Delta \gamma_{\sts\mb{p}} = \gamma_K -
\gamma_F$ can be expressed in terms of the moments of the distribution
of the hydrophobic residues on the recognition sites of the target and
the rival molecules, respectively:
\begin{equation}
  \Delta \gamma_{\sts\mb{p}} \sim - \frac{B\left([h^2_{\sts\mb{t}}] -
  [h_{\sts\mb{t}}][h_{\sts\mb{r}}]\right)}{2\left(C_2[h^2_{\sts\mb{t}}]
  - C_1 [h_{\sts\mb{r}}]\right)}  + \frac{B}{2C_2}
\end{equation}
Note that $C_1$ depends on $[h_{\sts\mb{t}}]$. For the special case
where the two types of molecules exhibit the same distribution one has
$[h_{\sts\mb{t}}] = [h_{\sts\mb{r}}] = [h]$. The shift is then
dominated by $\Delta \gamma_{\sts\mb{p}} \sim [h]^2$ in the asymptotic
limit of small values of the hydrophobicity $[h]$.  Assuming a linear
relation between the correlation length $\lambda_{\sts\mb{p}}$ and the
correlation parameter $\gamma_{\sts\mb{p}}$ in the parameter range
where the shift appears --- an assumption which should be valid if the
shift is small --- one also has $\Delta \lambda_{\sts\mb{p}} \sim
[h]^2$. The numerical Monte Carlo data in figure
\ref{bild:deltafreie_korrel} are consistent with this observation,
although it should be stressed that the quality of the shown numerical
data is not good enough to deduce reliable quantitative statements.

The mean field treatment has been used in this section to get an
expression for the dependence of the shift of the optimum correlation
lengths for the complementarity and the selectivity as a function of
the hydrophobicity of the target and rival molecules, respectively. To
this end, an expansion in the correlation parameter
$\gamma_{\sts\mb{p}}$ had been carried out, subsequently an average
over the correlated target and rival molecules was performed. The
coefficients of the series in $\gamma_{\sts\mb{p}}$ therefore
basically depend on the moments of the hydrophobicity distribution of
these molecules. It has to be noted in this context that the power
series in $\gamma_{\sts\mb{p}}$ is only an asymptotic one as for the
limit $\gamma_{\sts\mb{p}} \to 0$ the Hubbard-Stratonovich
transformation (\ref{eq:hubbard}) cannot be applied.  Nevertheless,
the mean field treatment gives reasonable results for the system with
correlated target and rival molecules as the optima of the
complementarity and the selectivity show up at non-zero values of the
correlation parameter $\gamma_{\sts\mb{p}}$. In the case of
uncorrelated target and rival molecules, however, this is not the case
(compare figure \ref{bild:freie_unkorrel}) and thus the mean field
treatment in the discussed framework is not applicable.

\section{Model of dominant cooperativity}
\label{kap:domkoop}

\begin{figure}[h!]
\begin{center}
\includegraphics[scale=0.285,angle=0]{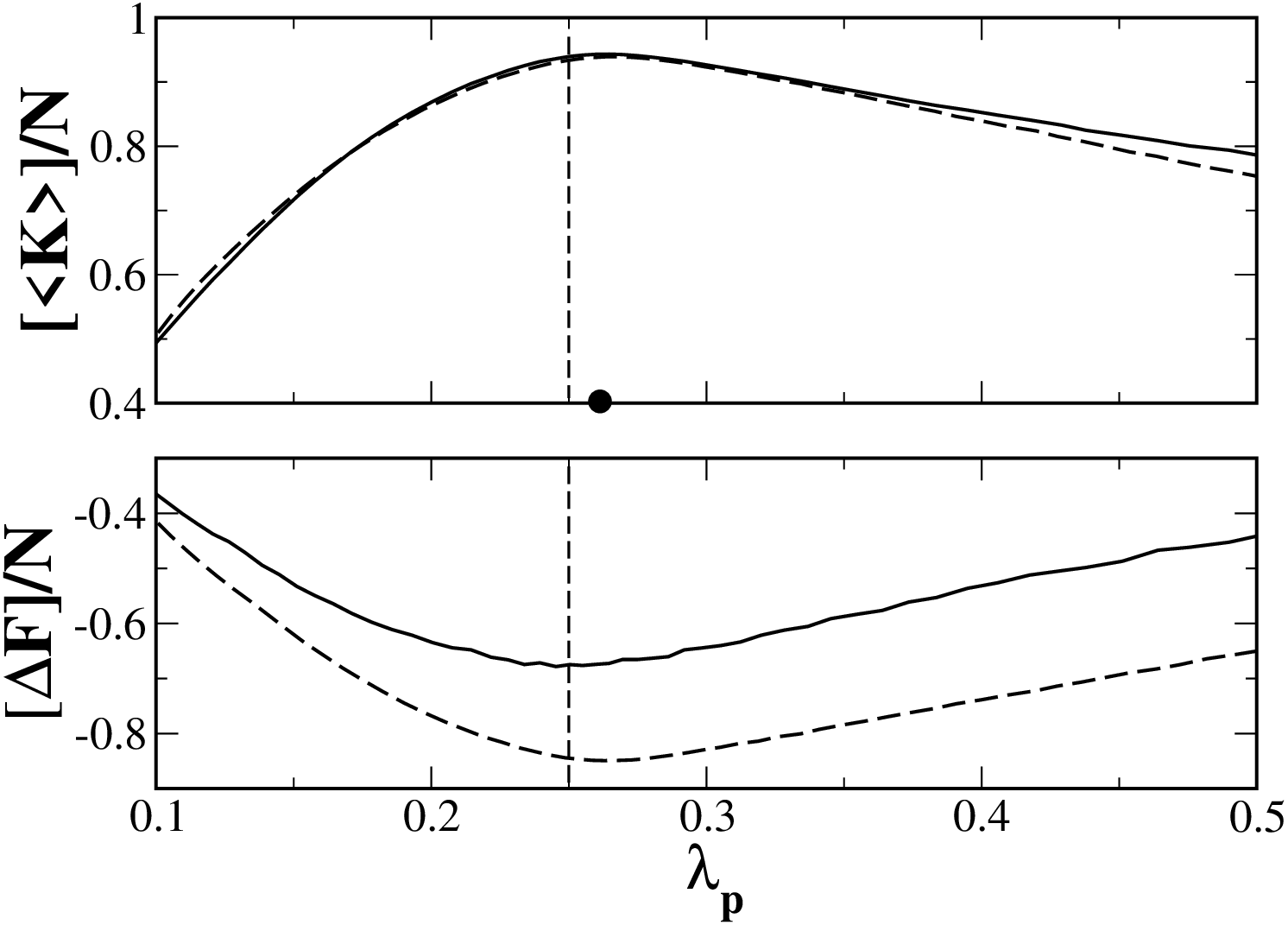}
\caption{\label{bild:freie_korrel_koopdom} The complementarity $\left
    [\left<K \right>\right]/N$ and the free energy difference $\left[
    \Delta F\right]/N$ of the system with dominant cooperative
  interactions as a function of the correlation length of the probe
  molecules. The correlation lengths of the target and rival molecules
  are fixed to the value shown by the circle, the corresponding
  hydrophobicities are $h_{\sts\mb{r}}=h_{\sts\mb{t}}=0.5$ (solid
  line) and $h_{\sts\mb{r}}=h_{\sts\mb{t}}=0.0$ (dashed line). The
  optimum correlation length for the recognition ability is clearly
  shifted to a value below the optimum value for the design of the
  probe ensemble for the interface with non-zero hydrophobicity. }
\end{center}
\end{figure}

In the previous section the constant $J$ of the cooperative
interaction term in (\ref{behringer:HP-coop}) has been set to zero so
that only the direct contact interactions due to the hydrophobic
effect contribute. In this section the influence of these additional
terms is taken into account. This is done by considering the case
where the cooperative interactions dominate over the direct contact
interactions. In \cite{Behringer_etal_2007b} it has been argued that
the Hamiltonian can be approximated by
\begin{equation}
\label{eq:hamil_domkoop}
\mathcal{H}_{\sts \mb{int}}(\sigma, \theta;s) = - \varepsilon\frac{1+s}{2}\sum_{i=1}^{N}
 \sigma_i \theta_i
\end{equation} 
in this case with the new (global) interaction variable $s$ taking on
the two possible values $\pm1$. Summing out the variable $s$ and
dropping irrelevant constants one ends up with the effective
Hamiltonian
\begin{equation}
  \mathcal{H}_{\sts \mb{eff}} = -\frac{\varepsilon}{2}\sum_i
  \sigma_i\theta_i - \frac{1}{\beta}\ln\cosh
  \left(\frac{\beta\varepsilon}{2} \sum_i\sigma_i\theta_i\right) 
\end{equation}
for the sequence $\theta$ of the probe molecule interacting with a
molecule whose sequence at its recognition site is specified by
$\sigma$.  Incorporating the correlation terms
(\ref{eq:correlhamilton}) the two stage approach to calculate the
recognition ability for a system with particular sequences for the
target and rival molecules can be carried out. The free energy
difference for the interaction of the probe molecules with the target
and the rival molecules, respectively, is then given by
\begin{eqnarray}
\label{eq:koopdom_freieener}
[\Delta F] &=& -\frac{\varepsilon}{2}\left[\left<\sum_i\sigma^{(\sts\mb{t})}_i\theta_i\right>_{P(\theta|\sigma^{(\sts\mb{t})})}\right]_{W^{(\sts\mb{t})}} +
  \frac{\varepsilon}{2} N [h_{\sts\mb{p}}]_{W^{(\sts\mb{t})}} h_{\sts\mb{r}} \\
&&-\frac{1}{\beta}\left[\left<\ln\cosh\left(\frac{\beta\varepsilon}{2}
  \sum_i\sigma_i^{(\sts\mb{t})}\theta_i\right)\right>_{P(\theta|\sigma^{(\sts\mb{t})})}\right]_{W^{(\sts\mb{t})}}
+\frac{1}{\beta}\left[\left<\ln\cosh
  \left(\frac{\beta\varepsilon}{2}
  \sum_i\sigma_i^{(\sts\mb{r})}\theta_i\right)\right>_{P(\theta|\sigma^{(\sts\mb{t})})}\right]_{W^{(\sts\mb{t})},W^{(\sts\mb{r})}}. 
\end{eqnarray}
The remaining averages in this expression of the free energy
difference can again be worked out by means of Monte Carlo
simulations. In figure \ref{bild:freie_korrel_koopdom} the
complementarity of the probe ensemble together with the free energy
difference is depicted as a function of the correlation length of the
probe molecules. Again the hydrophobicity of the target and rival
molecules is fixed, the hydrophobicity of the probe ensemble is
unrestricted (i.\,e. $\mu_{\sts \mb{p}} = 0$) and adjusts itself
during the design step. The data reveal again a shift in the optimum
correlation length for the recognition ability compared to the optimum
value for the complementarity, although this shift is somehow less
pronounced compared to the model with $J=0$.  Thus the findings for
the uncooperative model are reproduced qualitatively for the model
with additional cooperative interactions. Nevertheless a minor
difference is visible. Whereas the optimum correlation length with
respect to the complementarity of the probe molecules is clearly
shifted to a larger value compared to the fixed correlation length of
the target molecule in the case of the uncooperative model (compare
figure \ref{bild:comp_korrel}), the optimum appears (within the
accuracy of the numerics) at the same correlation length for the model
with dominant cooperativity. This is due to the fact, that the
cooperative interactions lead to the formation of extended patches of
good contacts \cite{Behringer_etal_2007b} and thus to an effective
reduction of the appearance of defects in the design step, which can
also be seen from the fact that the average complementarity at the
optimum correlation length is larger for the cooperative model (see
figures \ref{bild:comp_korrel} and \ref{bild:freie_korrel_koopdom}).
Thus defects need not be compensated by slightly extending the size of
the hydrophobic and polar patches due to correlation effects.

\begin{figure}[h!]
\begin{center}
\includegraphics[scale=0.285,angle=0]{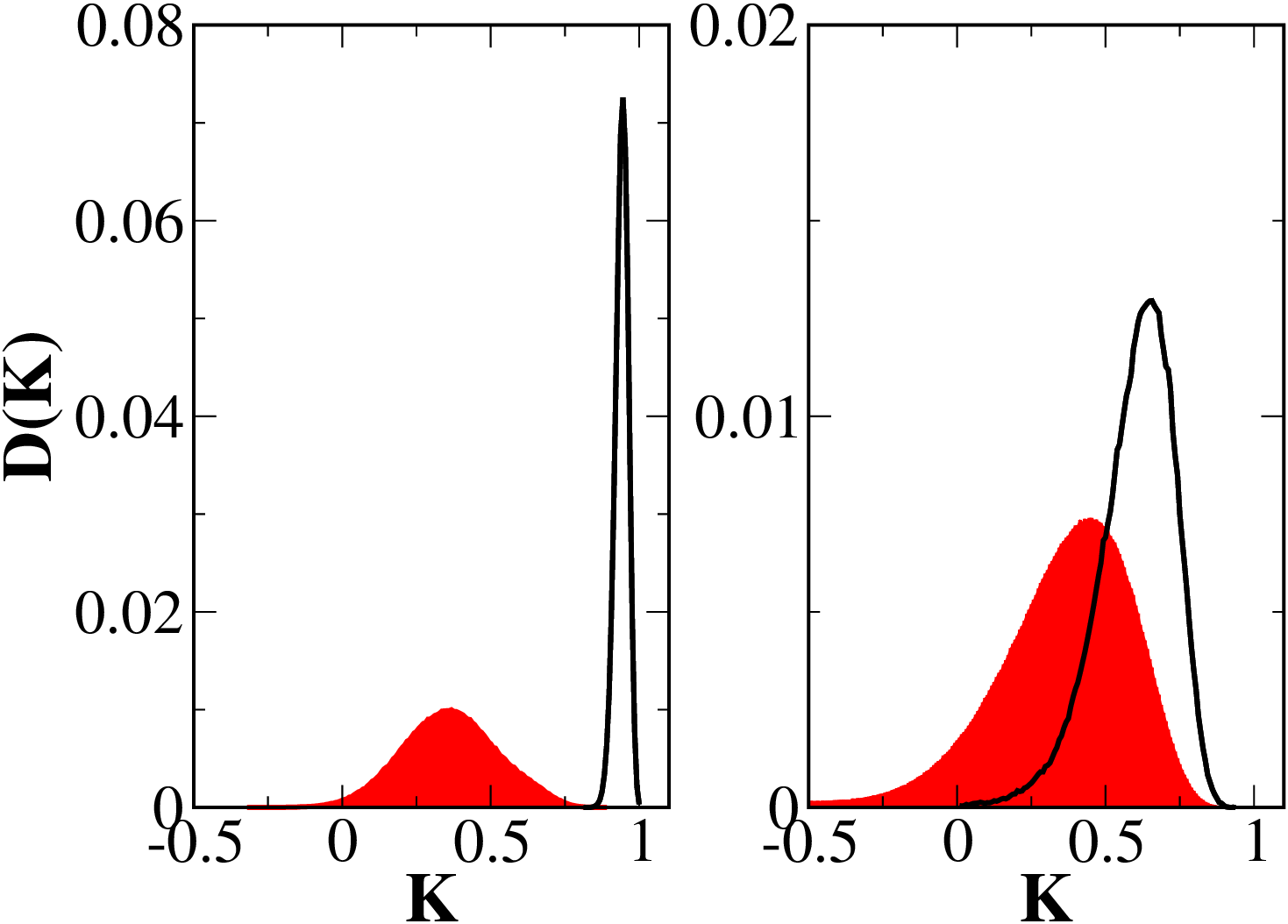}
\caption{\label{bild:kompdist_koopdom} Distribution of the
  complementarity of the probe ensemble with respect to the target
  molecules (solid line) and the rival molecules (shaded curve) for
  different correlation lengths on the recognition site of the probe
  molecules within the model of dominant cooperativity
  (\ref{eq:hamil_domkoop}). On the left hand side the correlation
  length $\lambda_{\sts\mb{p}} = 0.25$, on the right hand side
  $\lambda_{\sts\mb{p}} = 0.75$. The hydrophobicities are
  $h_{\sts\mb{t}}=h_{\sts\mb{r}}=0.4$, the correlation lengths of the
  target and rival molecules are
  $\lambda_{\sts\mb{t}}=\lambda_{\sts\mb{r}} = 0.263$ in each case.}
\end{center}
\end{figure}

As in the case of the uncooperative model (\ref{eq:h-unkooperativ})
the distribution function of the complementarity parameter of the
probe ensemble with respect to the target and rival molecules,
respectively, can be investigated. The corresponding curves in figure
\ref{bild:kompdist_koopdom} reveal that one ends up with the same
qualitative results as in the case of the uncooperative model. Note
that the two distributions are well separated from each other and that
the distribution of the complementarity with the target molecules is
fairly narrow for the correlation length that corresponds to a large
complementarity and selectivity. The width of the distribution of the
complementarity with the target is fairly reduced compared to the
width of the distribution for the uncooperative model (compare figure
\ref{bild:kompdist})

In principle the same numerical analysis of the recognition ability
can be carried out for arbitrary values of the cooperative interaction
constant $J$ in (\ref{behringer:HP-coop}) although in this case an
expression like (\ref{eq:koopdom_freieener}) for the free energy can
not be worked out and thus the numerical effort is much increased. The
free energy can be computed, for example, from the density of states
that can be evaluated by means of suitable Monte Carlo methods
\cite{Wang_2001,Hueller_2002,Landau_2004,Virnau_2004}.  As we expect
the qualitative physical behaviour not to change, we do not proceed
with such systems in this work.

\section{Summary and outlook}

In previous studies we developed coarse-grained lattice models to
analyse statistical properties of molecular recognition processes
between rigid biomolecules such as proteins
\cite{Behringer_etal_2006,Behringer_etal_2007,Behringer_etal_2007b}.
The general approach consists of two stages, where a design of probe
molecules with respect to a given target molecule is carried out
first. Afterwards the recognition ability of the probe molecules in an
heterogeneous environment with rival molecules is evaluated. Note that
the design step is carried out in absence of rival molecules whereas
the testing step includes rival molecules that are structurally
different from the target, but compete with them for the probe
molecules. In the present work we extended our previous models and
incorporated sequence correlations into our coarse-grained Hamiltonian
of the interactions across the interface of the two proteins. These
correlations affect the distribution of hydrophobic and polar residues
on the surfaces of the proteins. We investigated the extended models
by numerical Monte Carlo simulations and by mean field methods. Both
approaches lead to the same qualitative results. In particular we
computed the correlation length at which the optimum of the
complementarity of the design step appears.  The free energy
difference, that specifies the selectivity of the target-probe
interaction in the presence of rival molecules, shows an optimum at a
correlation length that is different from the one corresponding to the
optimum of the design step. This shift opens up the opportunity to
carry out the design slightly away from the optimum possible way
without losing selectivity. This might be relevant in the context of
harmful effects due to point mutations during evolution which our
design step is intended to mimic. In principle it should be possible
to check the appearance of two different correlation lengths for the
recognition sites of the two proteins that form a complex from
experimental structural data.  However, we do not know of a
corresponding study of this issue.

\acknowledgments

This work was funded by the Deutsche Forschungsgemeinschaft within the
collaborative research center SFB 613.

\end{document}